\documentclass[twocolumn,showpacs,prb,amsmath,amssymb,floatfix]{revtex4}
\usepackage{dcolumn}
\usepackage{amsmath,amssymb}
\usepackage{bm}
\usepackage{epsfig}
\usepackage{dsfont}

\begin{document}

\title{Effect of Coulomb interaction on the gap in monolayer and bilayer graphene}

\author{Andreas Sinner and Klaus Ziegler}

\vspace{2mm}  

\address{Institute for Physics, University of Augsburg, D-86135 Augsburg, Germany}


\date{\today}

\begin{abstract}
\noindent
We study effects of a repulsive Coulomb interaction on the spectral gap in monolayer and bilayer graphene in the vicinity of the charge neutrality point by employing the functional renormalization-group technique. In both cases Coulomb interaction supports the gap once it is open. For monolayer graphene we correctly reproduce results obtained previously by several authors, e.g., an apparent logarithmic divergence of the Fermi velocity and the gap as well as a fixed point corresponding to a quantum phase transition at infinitely large Coulomb interaction. On the other hand, we show that the gap introduces an additional length scale at which renormalization flow of diverging quantities saturates. An analogous analysis is also performed for bilayer graphene with similar results. We find an additional fixed point in the gapless regime with linear spectrum corresponding to the vanishing electronic band mass. This fixed point is unstable with respect to gap fluctuations and can not be reached as soon as the gap is opened. This preserves the quadratic scaling of the spectrum and finite electronic band mass. 

\end{abstract}
\maketitle

\section{Introduction}

Monolayer (ML) and bilayer (BL) graphene are semimetals with an electron and a hole band.
Both bands touch each other at two nodes. The low-energy dispersion in the
vicinity of these nodes is linear in ML graphene and quadratic in BL graphene. Exactly
at the nodes both systems obey a chiral symmetry which reflects the sublattice symmetry
of the underlying honeycomb lattice for ML or the inversion symmetry between single layers for BL. 
These symmetries can be broken, either by adding hydrogen atoms to ML graphene~\cite{FootNote}~[\onlinecite{Duplock2004, elias2009}],  
or by a biased gate voltage applied to BL graphene~[\onlinecite{ohta2006}]. The symmetry breaking is accompanied 
by opening of a gap in the spectrum. Then the question is whether or not such a gap is suppressed
or supported by the Coulomb interaction. Previous studies have shown that disorder 
induces random fluctuations of the gap which can suppress the effective gap and allow ML 
and BL graphene to be a conductor and to have a metal-insulator transition for a sufficiently
large average gap~[\onlinecite{Ziegler2009a,Ziegler2009b}]. On the other hand, it has been discussed 
that a short-range (Gross-Neveu) electron-electron interaction can dominate the long-range
Coulomb interaction, leading eventually to an insulating behavior~[\onlinecite{Juricic2009}]. 
In contrast to this works, we will follow subsequently a more direct route to an insulator by assuming
a small gap and study how this is affected by the Coulomb interaction itself.
The problem of Coulomb interaction in graphene has been previously studied by employing a perturbative 
renormalization-group (RG) approach, for clean ML 
graphene~[\onlinecite{Juricic2009,Son2007,Mishchenko2007,Mishchenko2008,Sheehy2007,Kane2006,Drut2009}] as well as for disordered ML graphene~[\onlinecite{Stauber2005,Foster2008,Herbut2008}]. 
These studies show clearly a strong renormalization of the Fermi velocity in the clean case 
and considerable interplay between Coulomb interaction and disorder. 

The outline of this paper is as follows. In Sec.~\ref{sec:model} we define the effective field theory for both graphene configurations with Coulomb interaction. We introduce the gap into the action by hand and perform decoupling in the interaction channel by means of the Hubbard-Stratonovich transformation. We obtain expressions for the bare bosonic and fermionic propagators and interaction vertices. In Sec.~\ref{sec:FRG} we write down renormalization-group flow equations for the gap parameter and fermionic wave-function renormalization factors and solve them for both gapless and gapped regimes. Furthermore, we analyze the fixed points for both graphene configurations. 

\section{The Model}
\label{sec:model}

We start with the zero temperature model for gapped ML and BL graphene. In the real space the Euclidean action of noninteracting ML and BL graphene in vicinity of a nodal point is given by 
\begin{eqnarray}
\label{eq:Model} 
{\cal S}^{}_0[\psi^\dag,\psi] &=& 
-\intop_{X}\psi^\dag_{X}\left(\hbar\partial^{}_t - i{\vec\tau}\cdot{\vec\nabla} + \Delta^{}_0\tau^{}_3\right)\psi^{}_X .
\end{eqnarray}
Here, Grassman fields $\psi^{\rm T}=(\psi^{}_{AK},\psi^{}_{BK},\psi^{}_{BK^\prime},\psi^{}_{AK^\prime})$ represent four-component spinors on the sublattices A and B in the vicinity of nodal points $K$ and $K^\prime$ in the momentum space which depend upon the $2+1$ dimensional vector $X$ that contains imaginary time $t$ and spatial vector $\vec x$ as components. Matrices $\tau^{}_{i,3}={\mathds 1}\otimes\sigma^{}_{i,3},\,i=1,2$, where $\sigma^{}_{i,3}$ denote usual Pauli matrices. 
For ML graphene the operator $\vec\nabla$ reads
\begin{equation}
{\vec\nabla} = \hbar v \partial^{}_{\vec x},
\end{equation}
where $v = \sqrt{3}ta/2\hbar$ denotes the bare (nonrenormalized) Fermi velocity and $\partial^{}_{\vec x}$ usual differential operators. For BL graphene it has the components:
\begin{eqnarray}
\nabla^{}_{1} &=& \frac{\hbar^2}{2\mu i}(\partial^{2}_{x^{}_1}-\partial^{2}_{x^{}_2}),\\
\nabla^{}_{2} &=& \frac{\hbar^2}{2\mu i} 2\partial^{}_{x^{}_1}\partial^{}_{x^{}_2},
\end{eqnarray}
with the bare band mass of electrons defined as 
$$
\mu = \frac{2t^{}_\perp\hbar^2}{3t^2a^2}.
$$
Here, $t$ and $t^{}_\perp$ are in- and out-of-plane hopping energies respectively; $a$ denotes the lattice spacing. The spectral gap $\Delta^{}_0$ is simply introduced by hand. 
The instantaneous interaction is the same for  both graphene configurations: 
\begin{equation}
{\cal S}^{}_c[\psi^\dag,\psi]= \frac{\hbar g}{2} \intop_{X}\intop_{X^\prime}(\psi^\dag\psi)^{}_{X}\frac{\delta(t-t^\prime)}{|\vec{x}-\vec{x}^\prime|}(\psi^\dag\psi)^{}_{X^\prime}.
\end{equation}
The microscopic strength of the Coulomb interaction between electrons is given by 
$$
g=\frac{e^2}{8\pi\epsilon^{}_0\epsilon\hbar} = \frac{\alpha c}{2\epsilon},
$$
where $e$ denotes the elementary charge, $\epsilon^{}_0$ the dielectric constant of the vacuum, $\alpha$ the fine structure constant, $c$ the speed of light in vacuum and $\epsilon$ the relative dielectric constant of the substrate. After performing a Fourier transform we obtain for both configurations 
\begin{eqnarray}
\nonumber
{\cal S}[\psi^\dag,\psi] &=& -\intop_Q{\psi^\dag_Q}\left[i\hbar q^{}_0 + {\vec h}\cdot{\vec\tau}+\Delta^{}_0\tau^{}_3\right]\psi^{}_Q \\
 \label{eq:ModelFourier} 
 &+&\frac{\hbar g^\prime}{2}\intop^{}_Q \frac{1}{q}~\rho^{}_{Q}\rho^{}_{-Q},
\end{eqnarray}
with different kinetic energy parts. The integrals over momentum and frequency $Q=(q^{}_0,{\vec q})$ 
with the absolute value of the momentum $q$ and zero-temperature Matsubara frequency $q^{}_0$ read 
$\intop_Q=(2\pi)^{-3}\int dq^{}_0d^2{\vec q}$ and should be thought of being regularized by means of an UV-cutoff $\Lambda^{}_0$. Furthermore we have re-scaled the interaction strength by the factor $2\pi$ that appears after Fourier transform, introducing $g^\prime=2\pi g$. The fermionic densities are defined as 
$$
\rho^{}_Q = \int^{}_P \psi^\dag_P\psi^{}_{P+Q}.
$$

For ML graphene the components of the vector ${\vec h}$ in the non-interacting part of the action read 
\begin{subequations}
\begin{equation}
\label{eq:ML-Hamiltonian}
h^{}_i = \hbar v q^{}_i,
\end{equation}
while for BL graphene 
\begin{equation}
\label{eq:BL-Hamiltonian}
h^{}_1 = \frac{\hbar^2}{2\mu}(q^{2}_1-q^{2}_2),\;\; h^{}_2=\frac{\hbar^2}{\mu} q^{}_1 q^{}_2.
\end{equation}
\end{subequations}
Below we will assume $\hbar=1$.

Now we map the pure fermionic action Eq.~(\ref{eq:ModelFourier}) onto the action containing both 
fermionic and bosonic degrees of freedom by means of the Hubbard - Stratonovich transformation~[\onlinecite{Schuetz2005}], 
\begin{eqnarray}
\label{eq:HubStrat}
{\cal S}^{}_0[\psi^\dag,\psi] + {\cal S}^{}_c[\psi^\dag,\psi] 
\to {\cal S}^{}_0[\psi^\dag,\psi] + {\cal S}^{}_0[\phi]+{\cal S}^{}_{Y}[\psi^\dag,\psi,\phi],
\end{eqnarray}
where the free bosonic action reads 
\begin{equation}
\label{eq:BosePart} 
{\cal S}^{}_0[\phi] = \frac{1}{2 g^\prime}\intop_Q q\phi^{}_Q\phi^{}_{-Q},
\end{equation}
and the third term denotes the interacting Yukawa term describing coupling between fermions and bosons:
\begin{equation}
\label{eq:Yukawa} 
{\cal S}^{}_{Y}[\psi^\dag,\psi,\phi] = i\intop_Q\intop_{K}\psi^\dag_{K}\psi^{}_{K+Q}\phi^{}_{-Q}.
\end{equation}

From the mixed action Eq.~(\ref{eq:HubStrat}) we obtain vertices and propagators by taking functional derivatives with respect to each field. 
From 
\begin{equation}
\left.\frac{\delta^2{\cal S}}{\delta\psi^{}_{Q}\delta\psi^\dag_{Q^\prime}}\right|_{\psi^\dag,\psi,\phi=0} =-(2\pi)^3\delta^{}_{Q,Q^\prime}G^{-1}_0(Q),
\end{equation}
we obtain the inverse fermionic propagator
\begin{equation}
\label{eq:BareFermiPropInv}
G^{-1}_0(Q) = iq^{}_0+{\vec h}\cdot{\vec\tau}+\Delta^{}_0\tau^{}_3, 
\end{equation}
with  components of the vector ${\vec h}$ defined in Eqs.~(\ref{eq:ML-Hamiltonian}) and (\ref{eq:BL-Hamiltonian}), and from 
\begin{equation}
\left.\frac{\delta^2{\cal S}}{\delta\phi^{}_{Q}\delta\phi^{}_{Q^\prime}}\right|_{\psi^\dag,\psi,\phi=0}=
-(2\pi)^3\delta^{}_{Q,-Q^\prime}F^{-1}(Q)
\end{equation}
the  inverse bosonic propagator
\begin{equation}
\label{eq:BareBosePropInv}
F^{-1}(Q) = -\frac{q}{g^\prime}. 
\end{equation}
Finally, the bare Yukawa vertex is obtained as 
\begin{equation}
\Gamma^{}(P^{}_1;P^{}_2,P^{}_3)=
\left.\frac{\delta^3{\cal S}}{\delta\psi^{}_{P_1}\delta\psi^{\dag}_{P_2}\delta\phi^{}_{P_3}}\right|_{\psi^\dag,\psi,\phi=0}.
\end{equation}
We arrive at
\begin{equation}
\label{eq:BareTriVert}
\Gamma^{}(P^{}_1;P^{}_2,P^{}_3) = i(2\pi)^3\delta^{}_{P^{}_1,P^{}_2+P^{}_3}.
\end{equation}

\section{Renormalization group equations}
\label{sec:FRG}

The functional RG is conveniently defined in terms of the field dependent functional of effective action ${\cal L}[\Phi]$, 
which in turn represents the Legendre transform of the generating functional of connected Green functions. For our purposes 
the ensemble average field $\Phi=(\bar\psi,\bar\psi^\dag,\bar\phi)$ is supposed to contain both fermionic and bosonic entries~[\onlinecite{Schuetz2005}]. 
The functional ${\cal L}$  depends on the IR-cutoff $\Lambda\leqslant\Lambda^{}_0$, which is eventually removed. 
The derivation of the functional RG flow equation is described in detail in Refs.~{[\onlinecite{Schuetz2005, Schuetz2006, Wetterich1993, Morris1994}]. The RG flow of ${\cal L}$ is generated by the regulator function introduced into the propagator of the non-interacting system   
and is determined by
\begin{equation}
\label{eq:FRGeq}
\partial^{}_{\Lambda}{\cal L}^{}_\Lambda[\Phi] = -\frac{1}{2}{\rm Tr}\left\{\partial^{}_\Lambda [G^{-1}_{0,R^{}_\Lambda}] \left(\frac{\delta^{2}{\cal L}^{R}_{\Lambda}}{\delta\Phi\delta\Phi}[\Phi]\right)^{-1}\right\},
\end{equation}
where $[G^{-1}_{0,R^{}_\Lambda}]$ is the propagator of the non-interacting system depending on the cutoff $\Lambda$ only via the regulator function. The matrix
\begin{equation}
\left.\frac{\delta^{2}{\cal L}^{R}_\Lambda}{\delta\Phi^{}_{}\delta\Phi^{}_{}}[\Phi]\right|^{}_{\Phi=0} = - [G^{R}_\Lambda]^{-1}=-([G^{-1}_{0,R^{}_\Lambda}]-\Sigma^{}_\Lambda)
\end{equation}
denotes the regularized full inverse propagator with $\Sigma^{}_\Lambda$ meaning the irreducible self-energy. 
The choice of the regulator will be specified a few lines below. Note that all quantities which appear on the right hand-side of Eq.~(\ref{eq:FRGeq}) dwell on the space of composite fields $\Phi$ and therefore represent 3$\times$3 matrices. 

Since our main interest is the determination of the spectrum renormalization of fermions due to the Coulomb interaction, we will focus on 
the coupling parameters in the fermionic sector of the theory. In the simplest approximation we make the following ansatz for the running effective action 
\begin{eqnarray}
\nonumber
{\cal L}^{}_\Lambda[\Phi] &\approx&
-\intop_Q \bar\psi^\dag_Q \left[iq^{}_0+Z^{-1}_\Lambda {\vec h}\cdot{\vec\tau}+\Delta^{}_\Lambda\tau^{}_3\right] \bar\psi^{}_Q\\
\nonumber
&&-\frac{1}{2}\intop_Q  F^{-1}(Q)\bar\phi^{}_Q\bar\phi^{}_{-Q}\\
\label{eq:TruncAct}
&&+i\intop_Q\intop_K\bar\psi^\dag_{Q}\bar\psi^{}_{Q+K}\bar\phi^{}_{-K},\;\;
\end{eqnarray}
which takes only the renormalization of the energy gap and of the electronic dispersion into account. 
We do not consider the renormalization of the Matsubara frequency since we assume the Coulomb 
interaction to be absolutely instantaneous. The inverse bosonic propagator $F^{-1}(Q)$ is defined in Eq.~(\ref{eq:BareBosePropInv}). 
For momenta larger than the UV-cutoff $\Lambda^{}_0$ the 
action in Eq.~(\ref{eq:TruncAct}) must reproduce the bare action from Eq.~(\ref{eq:HubStrat}). Therefore 
the initial conditions are chosen as $Z^{}_{\Lambda_0}=1$ and $\Delta^{}_{\Lambda^{}_0}=\Delta^{}_0$.

Taking functional derivatives with respect to both Grassmanian fields on both sides of Eq.~(\ref{eq:FRGeq}) and putting subsequently $\Phi=0$
we arrive at the RG flow equation for the inverse renormalized fermionic propagator. 
For details of its derivation we reffer to Refs.~[\onlinecite{Schuetz2005,Schuetz2006}]. 
If we employ the regularization scheme with the regulator built in the fermionic lines only, this equation can be written in the following algebraic form (note an additional minus sign due to Fermi statistics):
\begin{equation}
\label{eq:FermionicFlow}
\partial^{}_{\Lambda} G^{-1}_\Lambda(Q) = \intop^{}_P \dot{G}^{}_{\Lambda}(P) F(P-Q),
\end{equation}
where $F(Q)$ is the bare Coulomb potential defined in Eq.~(\ref{eq:BareBosePropInv}), and the single scale propagator $\dot{G}^{}_{\Lambda}$ is defined~as
\begin{equation}
\label{eq:SingScPr}
\dot{G}^{}_{\Lambda} = - G^{R}_\Lambda~\partial^{}_\Lambda [G^{-1}_{0,R^{}_\Lambda}]G^{R}_\Lambda
\end{equation}
We will work within the so-called sharp cutoff regularization scheme~[\onlinecite{Morris1994}]. Then the momentum cutoff is introduced as follows:
\begin{equation}
\label{eq:CutoffPropI}
G^{}_{0,R^{}_\Lambda}(Q)=\Theta(\Lambda<q<\Lambda^{}_0)G^{}_0(Q),
\end{equation}
where $\Theta(\Lambda<q<\Lambda^{}_0)=\Theta(\Lambda^{}_0-q)-\Theta(\Lambda-q)\to\Theta(q-\Lambda)$ as $\Lambda^{}_0\to\infty$. For momenta smaller than the UV-cutoff $\Lambda_0$, the flowing fermionic propagator $G^{R}_\Lambda(Q)$ is
\begin{equation}
\label{eq:FermFullPr} 
G^{R}_\Lambda(Q) = -\Theta(q-\Lambda)\frac{iq^{}_0-Z^{-1}_\Lambda {\vec h}\cdot{\vec\tau}-\Delta^{}_\Lambda\tau^{}_3}
{q^{2}_0 + E^2_\Lambda(q)},
\end{equation}
\noindent
and hence the single-scale propagator~[\onlinecite{Morris1994}] 
\begin{equation}
\label{eq:FermFullSSPr} 
\dot{G}^{}_\Lambda(Q) = \delta(q-\Lambda)\frac{iq^{}_0-Z^{-1}_\Lambda {\vec h}\cdot{\vec\tau}-\Delta^{}_\Lambda\tau^{}_3}
{q^{2}_0 + E^2_\Lambda(q)},
\end{equation}
where we have introduced $E^{}_\Lambda(q)=\displaystyle\sqrt{\Delta^2_\Lambda+\epsilon^2_\Lambda(q)}$ 
with the renormalized spectra of free fermions $\epsilon^{}_\Lambda(q) = Z^{-1}_\Lambda v q$ 
for ML and $\epsilon^{}_\Lambda(q)= {(2\mu Z^{}_\Lambda)^{-1} q^2}$ for BL. 
For both ML and BL the flow equations for the coupling parameter $\Delta^{}_\Lambda$ is extracted from Eq.~(\ref{eq:FermionicFlow}) in the same way:
\begin{subequations}
\begin{eqnarray}
\label{eq:EqDelta}
\partial^{}_\Lambda\Delta^{}_\Lambda &=& 
\frac{1}{4}{\rm Tr}^{}_{2}\left.\left\{\tau^{}_3\partial^{}_\Lambda G^{-1}_\Lambda(Q)\right\}\right|_{Q=0},
\end{eqnarray}
where ${\rm Tr}^{}_{2}$ denotes a trace operator acting on the pseudospin and valley space only. The RG flow equations for the factor $Z^{}_\Lambda$ are extracted differently for ML and BL due to the different scaling of the spectra in these configurations:
\begin{eqnarray}
\label{eq:EqZml}
{\rm ML}: &&
\partial^{}_\Lambda Z^{-1}_{\Lambda} = 
\frac{1}{4v}\frac{\partial}{\partial q^{}_i}{\rm Tr}^{}_{2}\left.\left\{\tau^{}_i \partial^{}_\Lambda G^{-1}_\Lambda(Q)\right\}\right|_{Q=0},\;\;\;\;\\
\label{eq:EqZbl}
{\rm BL}: &&
\partial^{}_\Lambda Z^{-1}_{\Lambda} = 
\frac{\mu}{4}
\frac{\partial^2}{\partial q^2_1}{\rm Tr}^{}_{2}\left.\left\{\tau^{}_1 \partial^{}_\Lambda G^{-1}_\Lambda(Q)\right\}\right|_{Q=0},
\end{eqnarray}
\end{subequations}
for $i=1,2$~\cite{footnote2}. Introducing the logarithmic flow parameter $\ell~=\log( \Lambda^{}_0/\Lambda)$ we obtain the same flow equation for the gap for both graphene configurations
\begin{subequations}
\begin{eqnarray}
\label{eq:Gap}
\partial^{}_\ell\Delta^{}_\ell &=& \frac{\bar{g}\Delta^{}_\ell\Lambda}{\sqrt{\Delta^2_\ell+\epsilon^{2}_\ell}},
\end{eqnarray}
where $\bar{g}=g^\prime/4\pi$, but different equations for the wave-function renormalization factor:
\begin{eqnarray}
\label{eq:ZetML} 
{\rm ML}:&&\partial^{}_\ell Z^{}_\ell = -\frac{1}{2}\frac{\bar{g} Z^{}_\ell\Lambda}{\sqrt{\Delta^2_\ell+\epsilon^2_\ell}},\\
\label{eq:ZetBL} 
{\rm BL}:&&\partial^{}_\ell Z^{}_\ell = -\frac{3}{8}\frac{\bar{g} Z^{}_\ell\Lambda}{\sqrt{\Delta^2_\ell+\epsilon^2_\ell}}.
\end{eqnarray} 
\end{subequations}
Here we have used the identity $\partial^{}_\ell Z^{-1}_\ell = -Z^{-2}_\ell\partial^{}_\ell Z^{}_\ell$. 
The flowing free fermion spectra are 
\begin{subequations}
\begin{eqnarray}
\label{eq:FlSpML}
{\rm ML}: && \epsilon^{}_\ell = v Z^{-1}_\ell\Lambda,\\
\label{eq:FlSpBL}
{\rm BL}: && \epsilon^{}_\ell = (2\mu Z^{}_\ell)^{-1}\Lambda^2.
\end{eqnarray}
\end{subequations}
The scaling dimension of the energy, (i.e. the dynamical exponent) is defined as $z=1-\eta^{}_\ell$ for ML and $z=2-\eta^{}_\ell$ for 
BL. Here, $\eta^{}_\ell$ is referred to as the anomalous dimension which can be obtained from the parameter $Z^{}_\ell$ by 
\begin{equation}
\label{eq:AnDimension}
\eta^{}_\ell = -\partial^{}_\ell \log Z^{}_\ell.
\end{equation}
Below we discuss solutions of these equations in gapless and gapped regimes.

\subsection{Gapless regime}

In the gapless regime Eqs.~(\ref{eq:Gap})-(\ref{eq:ZetBL}) are easily solved. The only solution of Eq.~(\ref{eq:Gap}) is the trivial one $\Delta^{}_\ell = \Delta^{}_{\ell=0}=0$, while Eqs.~(\ref{eq:ZetML}) and (\ref{eq:ZetBL}) reduce to 
$$
\partial^{}_\ell Z^{-1}_\ell~=~\lambda^{}_{\rm ML}
$$
with $\lambda^{}_{\rm ML}=\bar{g}/(2v)$ 
for ML and correspondingly 
$$
\partial^{}_\ell Z^{-1}_\ell=\lambda^{}_{\rm BL}e^\ell
$$
with $\lambda^{}_{\rm BL}=3\mu \bar{g}/(4 \Lambda^{}_0)$ for BL with solutions:
\begin{eqnarray}
\label{eq:SolZml} 
{\rm ML}: && Z^{-1}_\ell = 1 + \lambda^{}_{\rm ML}\ell,\\
\label{eq:SolZbl} 
{\rm BL}: && Z^{-1}_\ell = 1-\lambda^{}_{\rm BL}+\lambda^{}_{\rm BL}e^\ell.
\end{eqnarray}
The result of Eq.~(\ref{eq:SolZml}) corresponds to the well-known logarithmic renormalization of the Fermi velocity  $v^{}_\ell~=~Z^{-1}_\ell v$ in clean ML graphene due to the Coulomb interaction~[\onlinecite{Glonzalez1999,Mishchenko2007,Mishchenko2008,Son2007,Sheehy2007}]. 
Similarly, Eq.~(\ref{eq:SolZbl}) describes the renormalization of the electronic band mass $\mu^{}_\ell=Z^{}_\ell\mu$ in BL. At small momenta the band mass decreases proportionally to the momentum $\mu^{}_\ell\propto\Lambda$, i.e. the particles become effectively faster in analogy to ML.
Using Eq.~(\ref{eq:AnDimension}) we obtain expressions for the anomalous dimension 
\begin{subequations}
\begin{eqnarray} 
\label{eq:EtaML}
{\rm ML}: && \eta^{}_\ell = \lambda^{}_{\rm ML}Z^{}_\ell,\\
\label{eq:EtaBL}
{\rm BL}: && \eta^{}_\ell = \lambda^{}_{\rm BL} Z^{}_\ell e^\ell.
\end{eqnarray}
\end{subequations}
For $\ell\to0$ Eq.~(\ref{eq:EtaML}) approaches zero, meaning that the scaling dimension of the energy in ML remains $z=1$ and nothing changes the relativistic behavior of electrons. 
In contrast, Eq.~(\ref{eq:EtaBL}) approaches in this limit unity. This means that the scaling dimension of the energy $z=2-\eta^{}_\ell$ becomes  relativistic in BL with the velocity $v^{}_s=3\bar{g}/8$, i.e. in vacuum $c/v^{}_s\approx 1450$. Therefore, in absence of a gap in the spectrum of BL the Coulomb interaction attempts to linearize the fermionic dispersion in vicinity of the nodal points. Similar conclusions have been recently made by Kusminskiy et al.~[\onlinecite{CastroNeto2009}] for finite values of chemical potential. 
Their findings provided a good explanation of recent cyclotron experiments~[\onlinecite{Stormer2008}], where the effects discussed here have been observed. 

An estimation for the suitable scale below which this effect is observable can be made as follows: The only scale which affects the flow of the band mass in gapless BL graphene can be read off from Eq.~(\ref{eq:SolZbl}) (cf. Fig.~\ref{fig:BandMassBL}): 
\begin{equation}
\label{eq:ScaleBL}
\ell^\prime \approx \log\left(\frac{1-\lambda^{}_{\rm BL}}{\lambda^{}_{\rm BL}}\right). 
\end{equation}
Choosing $\Lambda^{}_0$ to be equal to the inverse lattice spacing we find for a realistic experimental situation ($\epsilon=1\div4$) $\ell^\prime\approx2.3\div3.8$ and the corresponding momentum scale  to be of the order 
$k^{}_c = \Lambda^{}_0e^{-\ell^\prime} \approx 1\cdot10^{-2}\div7\cdot10^{-2}~{\buildrel _{\circ}\over{\mathrm A}}^{-1}$. 

\subsection{Gapped regime}

\begin{figure}[t]
\includegraphics[height=4cm,width=7cm]{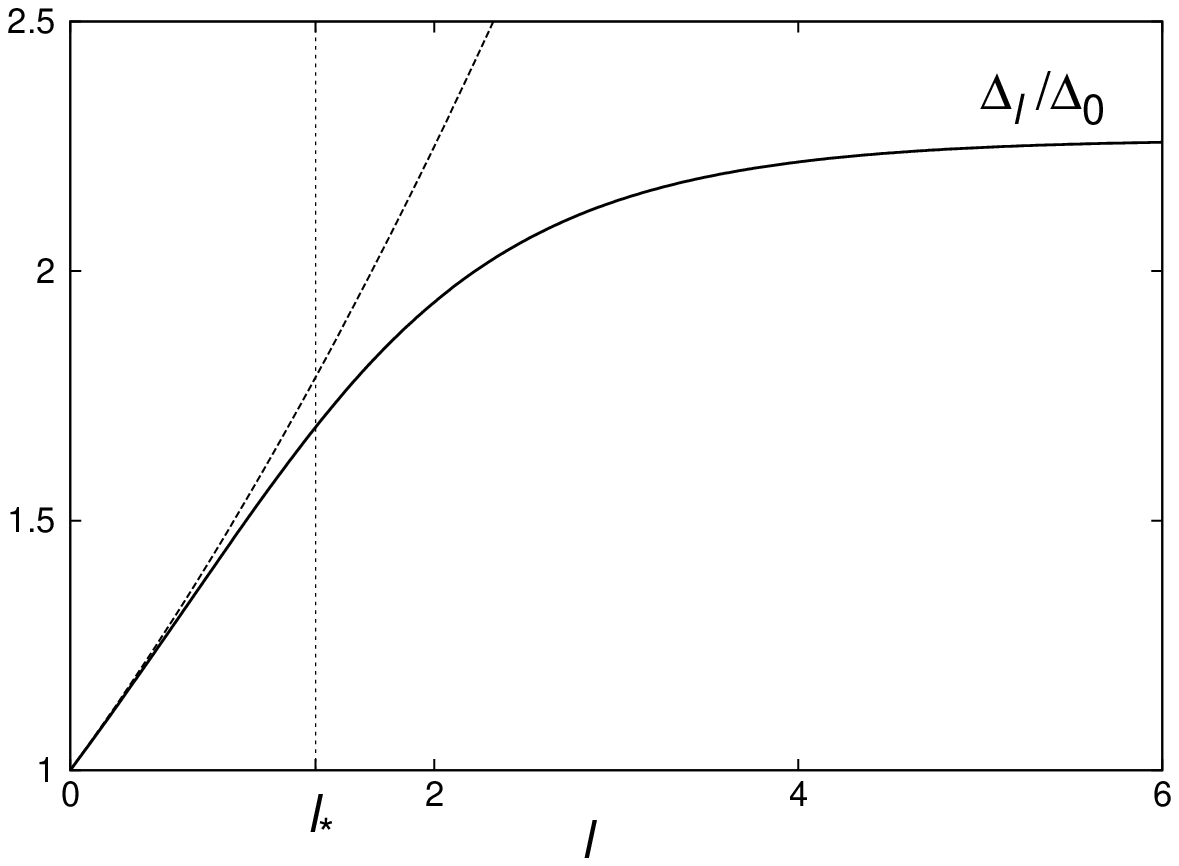}  
\caption{Renormalization of the gap $\Delta^{}_\ell$ in ML because of Coulomb interaction. 
The crossover scale $\ell^{}_\ast$ is determined from Eq.~(\ref{eq:ScaleEq}). 
The initial value of the gap is $\Delta^{}_0=0.2v^{}_0\Lambda^{}_0$, the dielectric constant $\epsilon=1$. The dashed line shows the Kane/Mele asymptote from Eq.~(\ref{eq:Kane}). The crossover scale $\ell^{}_\ast$ is determined from Eq.~(\ref{eq:ScaleEq}).} 
\label{fig:MLGap}
\includegraphics[height=4cm,width=7cm]{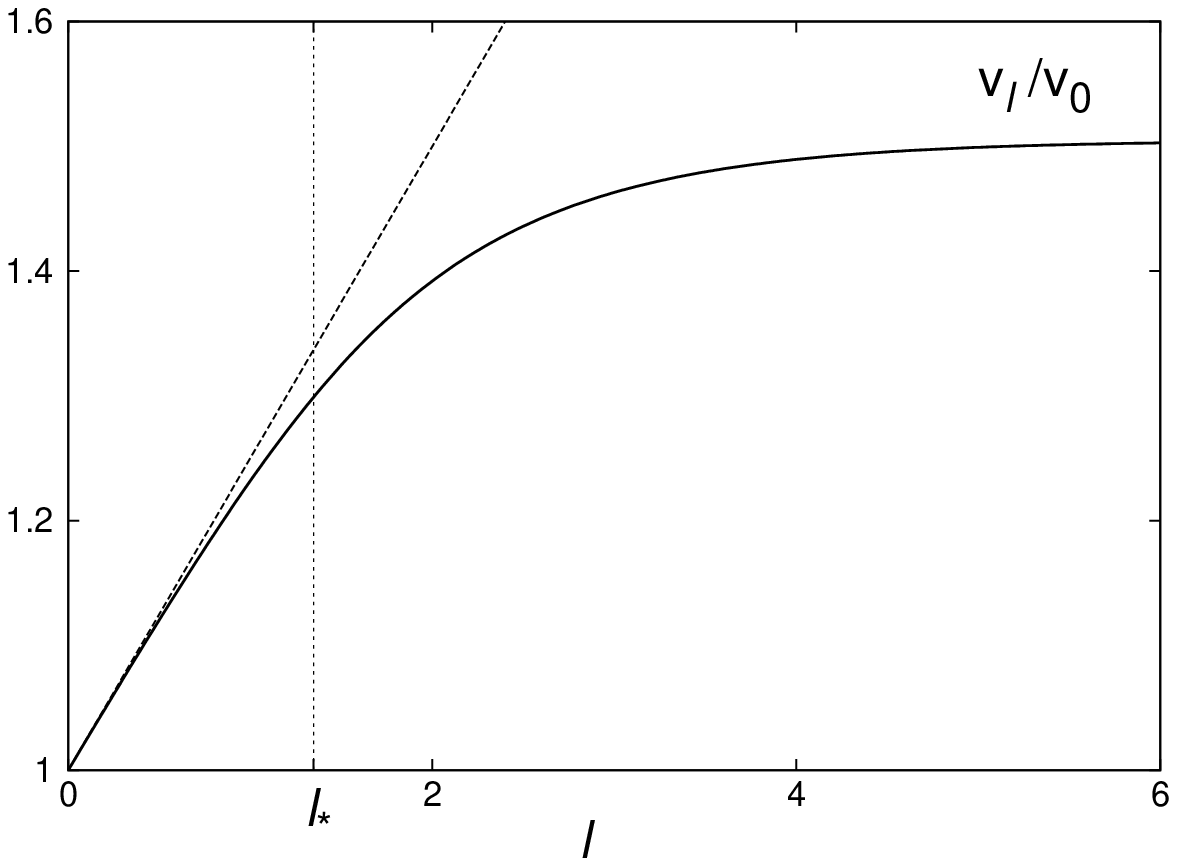}  
\caption{Renormalization of the Fermi velocity $v^{}_\ell=v^{}_0 Z^{-1}_\ell$ in ML because of Coulomb interaction with (solid line, full solutions of Eqs.~(\ref{eq:Gap}) and (\ref{eq:ZetML})) and without a gap (dashed line, Eq.~(\ref{eq:SolZml})). 
The crossover scale $\ell^{}_\ast$ is determined from Eq.~(\ref{eq:ScaleEq}).} 
\label{fig:FermiVelocity}
\end{figure}

Naively, for $\Delta^{}_\ell\ll\epsilon^{}_\ell$ we can neglect $\Delta^{}_\ell$ in the denominator. For ML we obtain from Eqs.~(\ref{eq:Gap}) 
\begin{eqnarray}
\label{eq:DlAs1ML}
\partial^{}_\ell\log\Delta^{}_\ell = 2\lambda^{}_{\rm ML}Z^{}_\ell,
\end{eqnarray}
which together with Eq.~(\ref{eq:SolZml}) 
reproduces the Kane/Mele result~[\onlinecite{Kane2006}]:
\begin{equation}
\label{eq:Kane}
\Delta^{}_\ell = \Delta^{}_0(1+\lambda^{}_{\rm ML}\ell)^2 = \Delta^{}_0 Z^{-2}_\ell,
\end{equation}
with the apparently logarithmically diverging gap. However, Eq.~(\ref{eq:Kane}) suggests that at large $\ell$ denominators in Eqs.~(\ref{eq:Gap}) and (\ref{eq:ZetML}) are dominated by the gap, i.e. $E^{}_{\ell\gg\ell^{}_\ast}=\sqrt{\Delta^2_\ell+\epsilon^{2}_\ell}\sim \Delta^{}_\ell$, 
where the crossover scale $\ell^{}_\ast$ can be determined from the condition 
\begin{equation}
\label{eq:ScaleEq}
\Delta^{}_{\ell^{}_\ast}\approx\epsilon^{}_{\ell^{}_\ast},
\end{equation}
which turns out to be a nonlinear algebraic equation if we take Eqs.~(\ref{eq:FlSpML}), (\ref{eq:SolZml}) and (\ref{eq:DlAs1ML}) into account. However, Eq.~(\ref{eq:ScaleEq}) can be uniquely solved numerically. The solution of Eq.~(\ref{eq:Gap}) in this case becomes
\begin{equation}
\label{eq:Dlrg}
\Delta^{}_\ell \approx \Delta^{}_{\ell^{}_\ast} + \bar{g}\Lambda^{}_\ast(1-e^{-\ell}),
\end{equation}
The physical gap is obtained for $\ell\to\infty$ 
\begin{equation}
\Delta^{}_c = \Delta^{}_{\ast}+{\bar{g}\Lambda^{}_\ast}\approx\epsilon^{}_{\ast}+{\bar g\Lambda^{}_\ast}.
\end{equation}
Therefore the Coulomb interaction in ML supports the gap once it is opened, independently of the bare gap magnitude. A typical flow of the gap parameter in ML is shown in Fig.~\ref{fig:MLGap}. At the same scale the logarithmic growth of the Fermi velocity stops and it also stabilizes at the finite value $v^{}_c \approx v\sqrt{1+\bar{g}\Lambda^{}_\ast/\Delta^{}_\ast}$ as depicted in Fig.~\ref{fig:FermiVelocity}.

\begin{figure}[t]
\includegraphics[height=4cm,width=7cm]{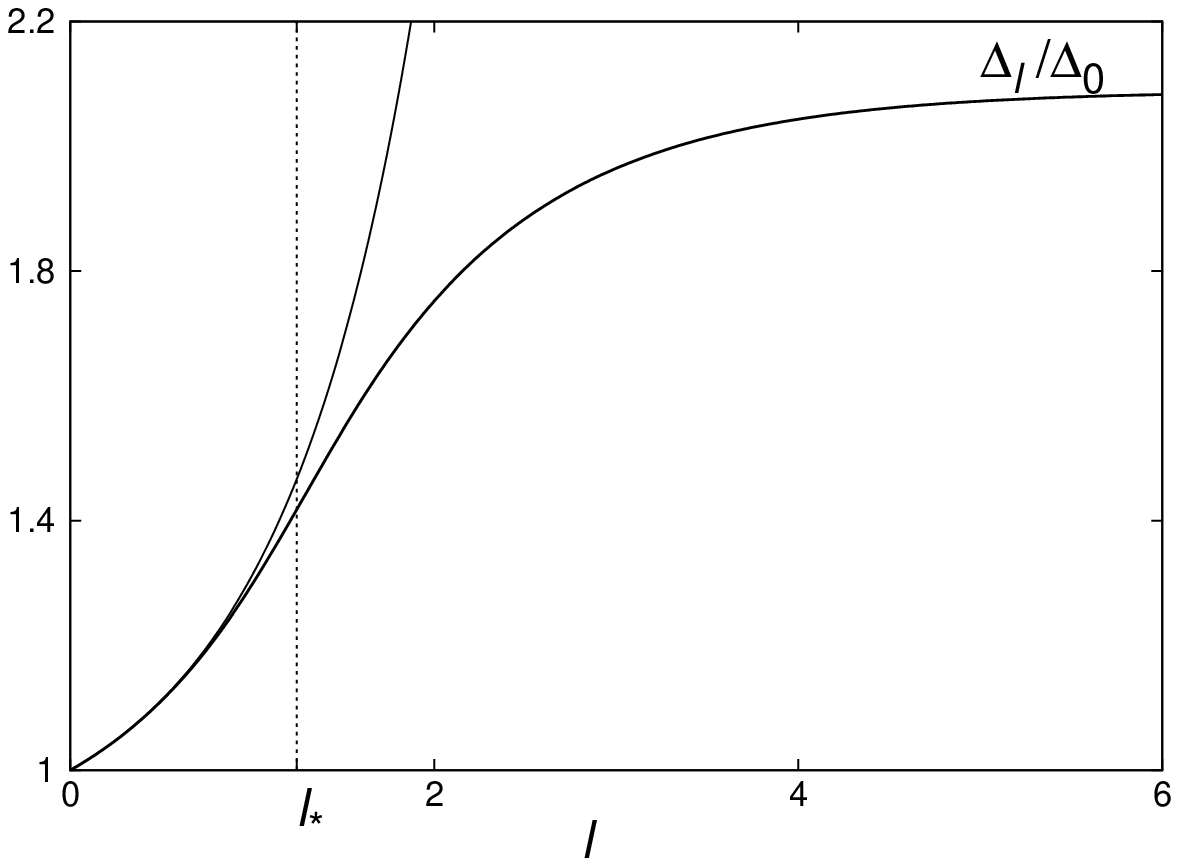}  
\caption{Renormalization of the gap $\Delta^{}_\ell$ in BL because of Coulomb interaction. The crossover scale $\ell^{}_\ast$ is determined from Eq.~(\ref{eq:ScaleEq}). The initial value of the gap is $\Delta^{}_0=0.2v^{}_0\Lambda^{}_0$, the dielectric constant $\epsilon=1$. The dashed line shows the large kinetic energy asymptote from Eq.~(\ref{eq:DlAs1BL}). The crossover scale $\ell^{}_\ast$ is determined from Eq.~(\ref{eq:ScaleEq}).} 
\label{fig:GapBLG}
\includegraphics[height=4cm,width=7cm]{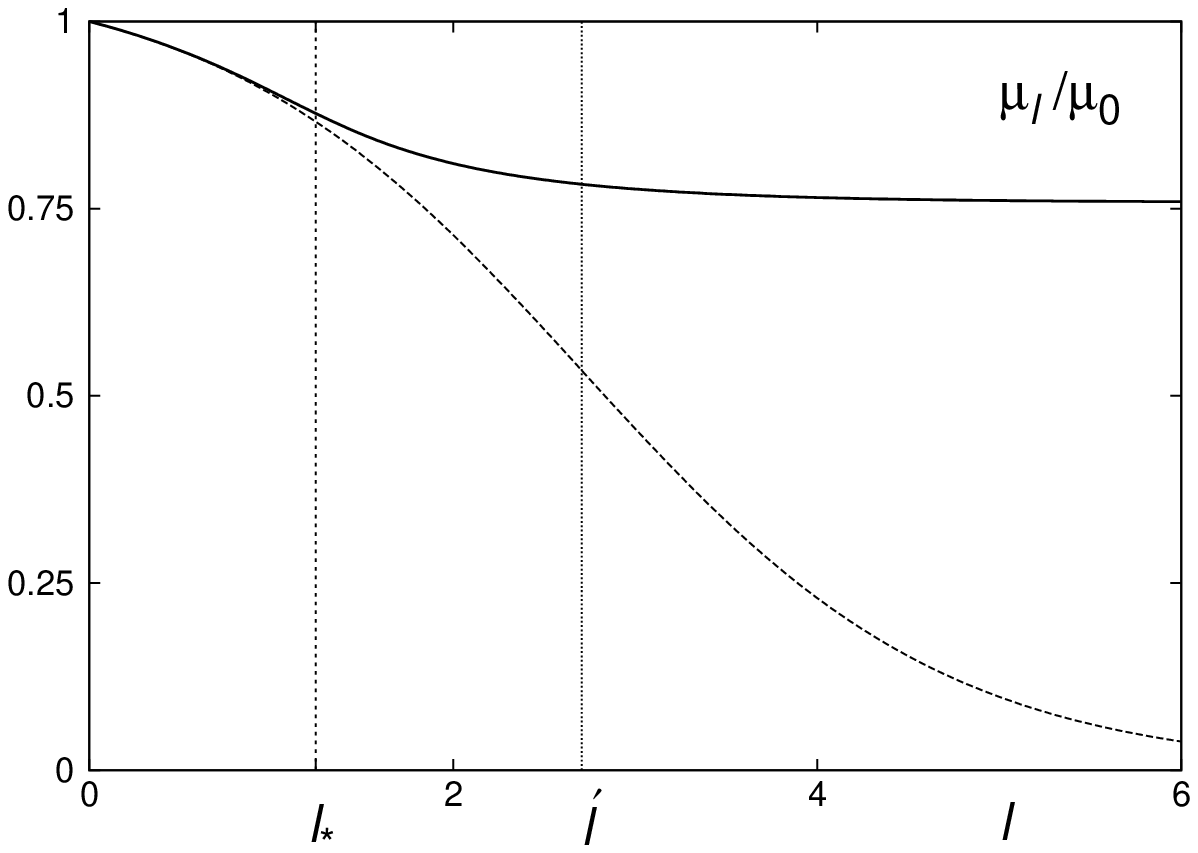}  
\caption{The renormalization of the band mass $\mu^{}_\ell = \mu^{}_0Z^{}_\ell$ due to the Coulomb interaction. The solid line shows the flow of the band mass of BL graphene with a gap. Dashed line shows asymptotic renormalization without a gap. In this case the electronic band mass scales to zero. This leads to the linear scaling of the spectrum. The scale $\ell^\prime$ is found from Eq.~(\ref{eq:ScaleBL}).} 
\label{fig:BandMassBL}
\end{figure}

The solutions for the gap in BL are similar in spirit but with an extra fixed point. For  $\Delta^{}_\ell\ll\epsilon^{}_\ell$ we obtain 
\begin{equation}
\label{eq:DlAs1BL}
\Delta^{}_\ell = \Delta^{}_0 Z^{-8/3}_\ell
\end{equation}
and therefore $\Delta_{\ell\to\infty}\to\infty$. The solution for $\Delta^{}_\ell\gg\epsilon^{}_\ell$ is formally given by Eq.~(\ref{eq:Dlrg}), too, such that the flow of the gap stabilizes at some finite value $\Delta^{}_\ast$ (cf. Fig.~\ref{fig:GapBLG}). On the other hand, the presence of the gap stabilizes the flow of the wave function renormalization factor $Z^{}_\ell$ and therefore the flow of the electronic band mass $\mu^{}_\ell = \mu^{}_0Z^{}_\ell$, which in this case remains finite (cf. Fig~\ref{fig:BandMassBL}). The scaling of the kinetic energy is in this case also preserved and remains equal to 2. 

In order to shed some light on the topological properties of the RG flow in the parametric space it is convenient to redefine Eqs.~(\ref{eq:ZetML}) and (\ref{eq:ZetBL}) in terms of kinetic energy and introduce dimensionless parameters by expressing both the gap and kinetic energy in units of  Coulomb energy:  
\begin{subequations}
\begin{eqnarray} 
\label{eq:RescGap}
\bar{\Delta}^{}_\ell &=& \frac{\Delta^{}_\ell}{\bar{g}\Lambda},\\
\label{eq:RescSpect}
\bar{\epsilon}^{}_\ell &=& \frac{\epsilon^{}_\ell}{\bar{g}\Lambda}, 
\end{eqnarray}
\end{subequations}
with $\epsilon^{}_\ell$ defined in Eq.~(\ref{eq:FlSpML}) for ML and in Eq.~(\ref{eq:FlSpML}) for BL. For both ML and BL we arrive at the same equation for the rescaled gap 
\begin{subequations}
\begin{eqnarray} 
\label{eq:GapFlowRescaled} 
\partial^{}_\ell\bar\Delta^{}_\ell = \displaystyle\bar\Delta^{}_\ell+\frac{\bar\Delta^{}_\ell}{\sqrt{\bar\epsilon^{2}_\ell+\bar\Delta^{2}_\ell}},
\end{eqnarray}
while equations for the rescaled kinetic energy are different due to different scaling behavior of spectra:
\begin{eqnarray} 
\label{eq:ZFlowRescaledMLG}
{\rm ML}: && \partial^{}_\ell\bar\epsilon^{}_\ell =  \displaystyle \frac{1}{2} \frac{\bar{\epsilon}^{}_\ell}{\sqrt{\bar{\epsilon}^2_\ell+\bar{\Delta}^2_\ell}},\\
\label{eq:ZFlowRescaledBLG}
{\rm BL}: && \partial^{}_\ell\bar\epsilon^{}_\ell = \displaystyle\frac{3}{8} \frac{\bar{\epsilon}^{}_\ell}{\sqrt{\bar{\epsilon}^2_\ell+\bar{\Delta}^2_\ell}} - \bar{\epsilon}^{}_\ell,
\end{eqnarray}
\end{subequations}
\begin{figure}[t]
\includegraphics[width=8cm]{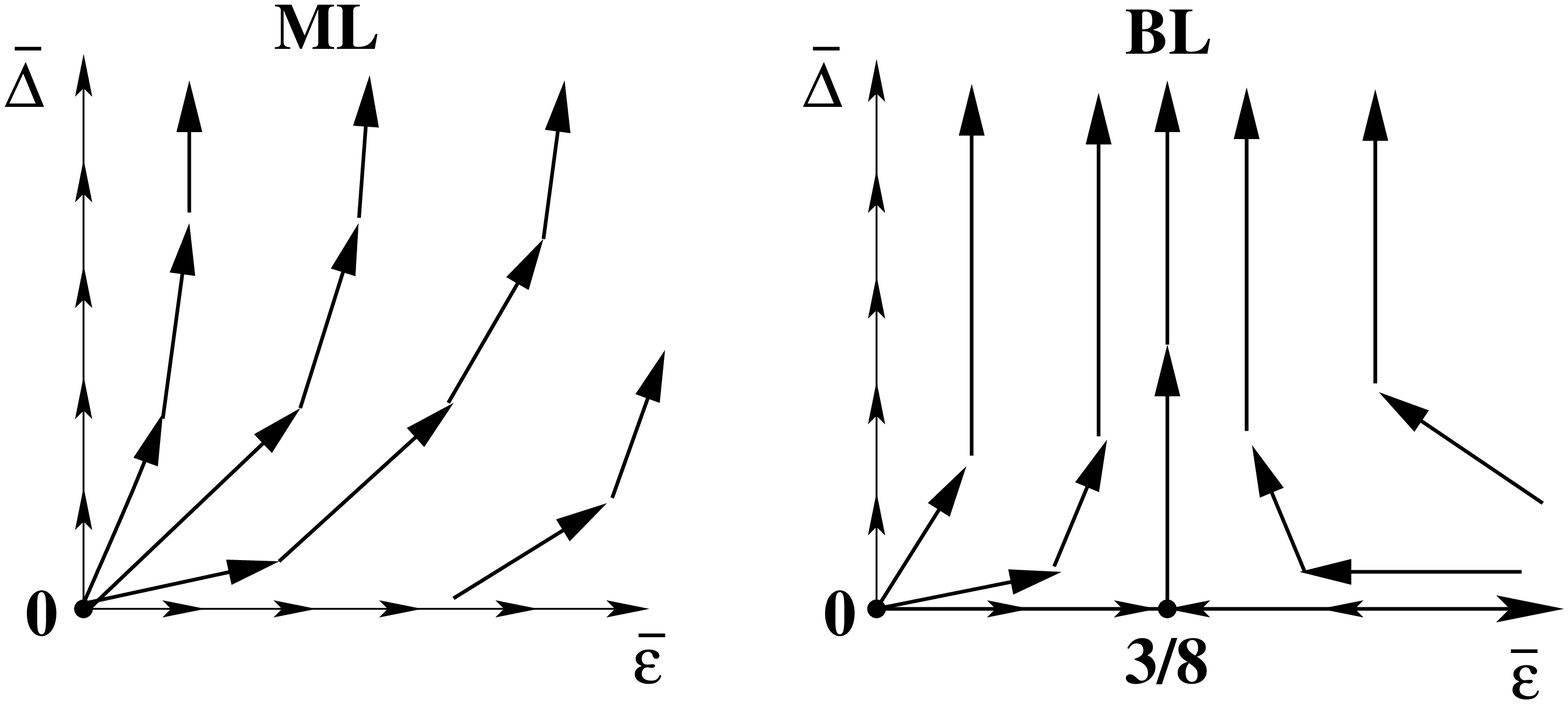}  
\caption{Schematic RG flow for both graphene configurations in the space spanned by the dimensionless kinetic energy $\bar\epsilon^{}_\ell$ and gap parameter $\bar\Delta^{}_\ell$.} 
\label{fig:FlowDiagMLG}
\end{figure}
The flow in the parametric space is schematically shown in 
Fig.~(\ref{fig:FlowDiagMLG}). The fixed points (FPs) 
are obtained by setting the right-hand sides of Eqs.~(\ref{eq:GapFlowRescaled})-(\ref{eq:ZFlowRescaledBLG}) 
to zero and solving the emerging system of algebraic equations. For ML graphene the only instable fixed point is at both $\bar\Delta^{}_\ell=0$ and $\bar\epsilon^{}_\ell=0$. From Eq.~(\ref{eq:RescSpect}) follows that this fixed point can be reached if 
\begin{equation}
\bar\epsilon^{}_\ell = \frac{vZ^{-1}_\ell}{\bar{g}}\to0.
\end{equation}
Since $Z^{-1}_\ell$ flows to a finite value, this can only be satisfied if $\bar{g}\to\infty$. This is a case of the famous quantum phase transition discussed in~Refs.~[\onlinecite{Son2007},\onlinecite{Sheehy2007}]. The instability of the fixed point means that the flow can leave it  in every direction. For any finite initial value of the gap it develops infinitely large value which indicates a finite physical gap. In contrast to the ML graphene, there is a nontrivial fixed point at finite dimensionless kinetic energy $\bar\epsilon^{}_\ast=3/8$ in the case of gapless BL graphene. This fixed point is characterized by the anomalous scaling dimension $\eta^{}_\ell=1$, i.e. the spectrum of BL becomes in this case linear. However this fixed point is instable with respect to the finite gap direction, i.e. once a small gap is opened in the spectrum the flow cannot reach this fixed point anymore but runs towards an infinite value. On the other hand, since the numerical value of $\bar\epsilon^{}_\ast$ at this fixed point suggests a strong coupling regime we might need to go beyond the truncation Eq.~(\ref{eq:TruncAct}) and take additionally flow of the $\bar\psi\bar\psi^\dag\bar\phi$--vertex into account.

\section{Conclusions}
\label{sec:conclusions}

In conclusion, we have studied both ML and BL graphene with Coulomb interaction and a uniform gap by employing a renormalization-group technique. 
In contrast to previous approaches to gapped ML graphene based on the renormalization group approach  [\onlinecite{Juricic2009,Mishchenko2007,Sheehy2007,Kane2006}], which predict logarithmically divergent renormalization of the gap and the Fermi velocity, our results suggest a saturation of RG flows at an intrinsic scale related to the gap. This saturation takes place for both ML and BL graphene, for any finite initial value of the gap no matter how small it is, and since measured quantities should be finite, this might be suggestive of a gap in the spectrum of both configurations at energies below 0.1 eV.

In ML graphene the Coulomb energy exhibits the same scaling as the kinetic energy. Once a spectral gap is opened it creates an additional length scale which dominates the physics at small momenta. This scale cuts off the logarithmic divergence of the Fermi velocity and gap itself such that the flow of both quantities stabilizes at the finite value.

For gapless BL graphene is shown that Coulomb interaction renormalizes the electronic band mass which scales to zero for small momenta. This leads to a paradoxical result that the electronic spectrum should become linear close to the charge neutrality point. This regime corresponds to a stable fixed point and therefore the flow should inevitably go into this point. 
The quadratic scaling of the spectrum is rescued by the presence of the gap, since for any finite starting values of the gap the flow of the band mass always saturates at a finite value.

\section*{ACKNOWLEDGEMENTS}
We gratefully acknowledge useful discussions with S.~Savel'ev, B.~D\'{o}ra, and A.~Sedrakyan. We thank A.~H.~Castro Neto for bringing Refs.~[\onlinecite{Stormer2008}] and [\onlinecite{CastroNeto2009}] to our attention. This work has been supported by the DPG-grant ZI 305/5-1.


\begin{thebibliography}{99}
\bibitem{FootNote} It should be mentioned that it is difficult to distinguish a 
genuine gap from the disorder generated localization near the nodal points in experimental situations.
\bibitem{footnote2}  The same flow equation for the wave renormalization factor $Z^{}_\Lambda$ for BL can be extracted in several ways because of  the different components of the vector $\vec h$.
\end{thebibliography}
\end{document}